\documentclass[a4paper, 12pt]{article}
\usepackage[T1]{fontenc}
\usepackage{jheppub,graphicx,epsf,amssymb,amsbsy,amsfonts,amssymb,amsmath, empheq,mathrsfs,appendix,soul}
\usepackage{hyperref}

\def\be{\begin{equation}}
\def\ee{\end{equation}}
\def\bea{\begin{eqnarray}}
\def\eea{\end{eqnarray}}

\def\<{\left<}
\def\>{\right>}
\def\bg{\bar{g}}
\def\beq{\begin{eqnarray}}\def\eeq{\end{eqnarray}}
\def\ba#1\ea{\begin{align}#1\end{align}}
\def\bg#1\eg{\begin{gather}#1\end{gather}}
\def\bm#1\em{\begin{multline}#1\end{multline}}
\def\bmd#1\emd{\begin{multlined}#1\end{multlined}}

\def\D{\Delta}

\def\({\left(}
\def\){\right)}
\def\[{\left[}
\def\]{\right]}

\def\D{\Delta}

\title{Holographic symmetry algebra for the MHV sector revisited}

\author[1,2]{Shamik Banerjee}
\author[1]{Mousumi Maitra}
\author[1,2]{,~Raju Mandal}
\author[3]{,~Milan Patra}
\affiliation[1]{National Institute of Science Education and Research (NISER),Bhubaneswar 752050, Odisha, India}
\affiliation[2]{Homi Bhabha National Institute, Training School Complex, Anushakti Nagar, Mumbai 400094, India}
\affiliation[3]{Centre for High Energy Physics, Indian Institute of Science, C.V. Raman Avenue, Bangalore 560012, India.}
\emailAdd{banerjeeshamik.phy@gmail.com, mousumimaitra84@gmail.com, rajuphys002@gmail.com, milanpatra1515@gmail.com}
\abstract{We revisit the holographic symmetry algebra in the MHV sector. We find an infinite dimensional Abelian symmetry algebra whose generators are the conformally soft negative helicity gravitons and gluons. So the complete symmetry algebra in the MHV graviton sector is a semideirect product of the $w_{1+\infty}$ algebra and the infinite dimensional Abelian algebra. Similarly in the MHV gluon sector the symmetry algebra is a semidirect product of the $S$ algebra and the infinite dimensional Abelian algebra. The extended symmetry algebra has some use. For example, it is known for sometime that an $n$ point MHV amplitude satisfies $(n-2)$ Knizhnik-Zamolodchikov (KZ) type equations. So two equations are missing. We show that the extended symmetry algebra has additional null states whose decoupling give rise to the two missing equations.} 
\begin{document}
\maketitle
\flushbottom

\section{Introduction}
Symmetry plays an important role in physics and identifying all the non-trivial symmetries of the S matrix in asymptotically flat spacetime(AFS) still remains an open problem. Celestial Holography \cite{gerger:2017zoo,Strominger:2013lka,Banerjee:2020zlg,Banerjee:2021cly,Guevara:2021abz,Strominger:2021lvk,Ball:2021tmb,Adamo:2021lrv,Guevara:2025tsm,Banerjee:2023zip,Banerjee:2020vnt,Banerjee:2023bni} is a very useful framework to address such issues. In this paper we revisit the soft symmetries or the Holographic Symmetries Algebra \cite{Guevara:2021abz,Strominger:2021lvk} for the (tree level) MHV scattering amplitudes. In \cite{Banerjee:2020zlg}, it was shown that the MHV graviton scattering amplitudes are completely determined by the symmetry algebra which is a semidirect product of a  $SL_2(\mathbb R)$ current algebra and supertranslations. It works in the following way. One can show that the representation of the symmetry algebra, which is relevant for the MHV amplitudes, has null states. These null states are similar to the Knizhnik-Zamolodchikov (KZ) null states that are there in the WZW models. The existence of these null states lead to differential equations for the Celestial correlators which one can solve to get the MHV amplitudes. However, there is a puzzle. For an $n$-point MHV amplitude one gets only $(n-2)$ KZ type differential equations. This is unlike the case of the WZW model where there are $n$ KZ equations for an $n$ point correlation function. This puzzle is not solved by the extended symmetry algebra \cite{Strominger:2021lvk} of the MHV sector which is $w_{1+\infty}$. 

In this paper we address this puzzle and get the other two missing equations. We show that the complete symmetry algebra of the MHV sector is a semidirect product of the $w_{1+\infty}$ algebra and an infinite Abelian algebra whose generators are the \textit{conformally soft negative} helicity gravitons in the MHV sector. This additional Abelian symmetry gives us the missing null states. Note that there is \textit{no} energetically soft \textit{negative} helicity graviton in the MHV sector. So our results also highlight the importance and usefulness of the concept of conformal softness \cite{Fan:2019emx, Pate:2019mfs,Donnay:2018neh,Nandan:2019jas,Adamo:2019ipt,Puhm:2019zbl} in the context of Celestial Holography and asymptotic symmetries. 

A similar story exists for the MHV gluon scattering amplitudes which we discuss in the detail in the paper.

The paper is organized as follows: We start with the mixed helicity celestial OPE between two graviton primary operators in the MHV-sector and write down the corresponding symmetry algebra in section \ref{algebra1}. We also review the $w_{1+\infty}$ algebra that is obtained from the OPE between two positive-helicity graviton primaries. In section \ref{gluon}, we discuss the symmetry algebra for MHV gluon scattering amplitudes. In section \ref{kz1}, we talk about new KZ-type null states for negative-helicity gravitons and we find the same for gluons in section \ref{kz2}.  Appendix \ref{graviton_a} and appendix \ref{gluon_b} contain the derivations of the new symmetry algebra from the mixed helicity OPE of the MHV-sector for gravitons and gluons respectively. The derivation of the KZ-type null states for negative-helicity gluons is in appendix \ref{kz_d}. We write down the conditions on the graviton primary operators for the symmetry algebra under consideration in appendix \ref{primaries}.
\section{Symmetry algebra from negative helicity graviton in the MHV-sector}\label{algebra1}
In this section we derive the additional symmetry generators in the MHV-sector which are obtained from a ``conformally soft'' negative helicity graviton. Let $G^{\pm}_{\Delta}(z,\bar{z})$ denote a positive/negative helicity outgoing graviton conformal-primary of dimension $\Delta$ at the point $(z,\bar{z})$ on the celestial torus.
We start with the mixed helicity OPE in the MHV sector\footnote{If we go beyond MHV sector then positive helicity graviton also appears in the RHS of \eqref{ope}.}. The singular terms are given by \cite{Guevara:2021abz} 

\be
\label{ope}
\begin{gathered}
G^-_{\D_1}(z_1,\bar z_1) G^+_{\D_2}(z_2,\bar z_2) \sim 
-\frac{\bar z_{12}}{z_{12}} \sum_{n=0}^{\infty} B(\D_1 +3 +n, \D_2 -1) \frac{(\bar z_{12})^n}{n!} \bar\partial_2^n G^-_{\D_1+\D_2}(z_2,\bar z_2)
\end{gathered}
\ee

Now we define an infinite family of conformally soft negative-helicity gravitons $\bar H^k(z,\bar z)$ as


\be\label{cs}
\bar H^k(z,\bar z) = \lim_{\D\rightarrow k} (\D-k) G^-_{\D}(z,\bar z),~~~k=-2,-3,-4,-5,\cdots 
\ee
with weights 
\begin{equation}
	\label{eq:weight1}
	(h,\bar{h})=\Bigg(\frac{k-2}{2},\frac{k+2}{2}\Bigg).
\end{equation}
Note that the limit \eqref{cs} is different from the ``energetic soft limit" because in the MHV sector energetic soft limit of negative helicity gravitons is identically zero.  

Now from \eqref{ope} we can see that the OPE truncates in $\bar z$ when $k=-2,-3,-4\cdots$. 
So we can do a truncated mode expansion of $\bar H^k(z,\bar z)$ as follows

\be
\bar H^k(z,\bar z) = \sum_{m= \frac{k+2}{2}}^{-\frac{k+2}{2}} \frac{\bar H^k_m(z)}{\bar z^{m+\frac{k+2}{2}}}
\ee

The currents $\bar H^k_m(z)$ admit the usual Laurent expansion in $z$

\be
\bar H^k_m(z) = \sum_{\alpha \in \mathbb{Z} - \frac{k-2}{2}} \frac{\bar H^{k}_{\alpha,m}}{z^{\alpha+\frac{k-2}{2}}}
\ee

Similarly, one defines an infinite tower of conformally soft positive-helicity graviton operators
\begin{equation}
	\begin{split}
		H^k\left(z,\bar{z}\right)=\lim_{\Delta\to k}\left(\Delta-k\right)G^+_{\Delta}\left(z,\bar{z}\right),~~~k=2,1,0,-1,\cdots
	\end{split}
\end{equation}
with weights
\begin{equation}
	\label{eq:weight2}
	(h,\bar{h})=\Bigg(\frac{k+2}{2},\frac{k-2}{2}\Bigg).
\end{equation}
Now it is well known that the modes of the currents $H^k(z,\bar z)$ satisfy the holographic symmetry algebra \cite{Guevara:2021abz}
%
\begin{equation}
		\label{eq:algebra01}
		\begin{split}
			\left[H^{k_{1}}_{\alpha_{1},m_{1}},H^{ k_{2}}_{\alpha_{2},m_{2}}\right]=&-\Big[
			m_{2}(2-k_{1})-m_{1}(2-k_{2}))
			\Big]\\
			&\times\frac{\Big(\frac{2-k_1}{2}-m_1+\frac{2-k_2}{2}-m_2-1\Big)!}
			      {\Big( \frac{2-k_1}{2}-m_1 \Big)!
			       \Big( \frac{2-k_2}{2}-m_2 \Big)!}
		     \frac{\Big(\frac{2-k_1}{2}+m_1+\frac{2-k_2}{2}+m_2-1\Big)!}
		     {\Big( \frac{2-k_1}{2}+m_1 \Big)!
		     	\Big( \frac{2-k_2}{2}+m_2 \Big)!}  
	     	H^{k_1+k_2}_{\alpha_1+\alpha_2,m_1+m_2}
		\end{split}
	\end{equation}
Similarly, the OPE \eqref{ope} gives the commutator algebra between the modes of $H^k(z,\bar z)$ and $\bar H^k(z,\bar z)$ to be  (the detail of the computation is in Appendix \ref{graviton_a})	
\begin{equation}
	\label{eq:algebra02}
	\begin{split}
		\left[H^{ k_{1}}_{\alpha_{1},m_{1}},\bar{H}^{ l_{1}}_{\beta_{1},n_{1}}\right]
		=&-\Big[
		     n_1(2-k_1)+m_1(2+l_1)
		   \Big]\\
		   &\times\frac{\Big(\frac{2-k_1}{2}-m_1-\frac{2+l_1}{2}-n_1-1\Big)!}
		        {\Big(\frac{2-k_1}{2}-m_1\Big)!
		         \Big(-\frac{2+l_1}{2}-n_1\Big)!}
	      \frac{\Big(\frac{2-k_1}{2}+m_1-\frac{2+l_1}{2}+n_1-1\Big)!}
	      {\Big(\frac{2-k_1}{2}+m_1\Big)!
	       \Big(-\frac{2+l_1}{2}+n_1\Big)!}
        \bar{H}^{k_1+l_1}_{\alpha_1+\beta_1,m_1+n_1}
	\end{split}
\end{equation}
Lastly, the OPE between two negative helicity gravitons does not have a pole term in the MHV sector and so we should have 
%

\begin{equation}
	\label{eq:algebra03}
	\left[\bar{H}^{ l_{1}}_{\beta_{1},n_{1}},\bar{H}^{ l_{2}}_{\beta_{2},n_{2}}\right]=0.
\end{equation}
Now we define the positive-helicity and negative-helicity light transformed soft graviton operators as \cite{Strominger:2021lvk}
\begin{equation}
	\label{eq:w1}
	w^{p}_{m}=\frac{1}{\kappa}(p-1-m)!(p-1+m)!H^{-2p+4}_{m}
\end{equation}
\begin{equation}
	\label{eq:w2}
	\bar{w}^{q}_{n}=\frac{1}{\kappa}(q-1-n)!(q-1+n)!\bar{H}^{-2q}_{n}
\end{equation}
where $p,q=1,\frac{3}{2},2,\frac{5}{2},...$ and $m$ and $n$  are restricted to be
\begin{equation}
	\label{eq:res1}
	1-p \le m \le p-1; \hspace{4mm}  1-q \le n \le q-1
\end{equation}
The algebra of these light transformed operators is given by
\begin{equation}\label{eq:w3}	
\begin{gathered}
	[w^{p}_{m},w^{q}_{n}]=[m(q-1)-n(p-1)]w^{p+q-2}_{m+n} \\
    [w^{p}_{m},\bar{w}^{q}_{n}]=[m(q-1)-n(p-1)]\bar{w}^{p+q-2}_{m+n} \\
    [\bar{w}^{p}_{m},\bar{w}^{q}_{n}]=0
    \end{gathered}
\end{equation}
The MHV graviton scattering amplitudes are governed by the above symmetry algebra. The conformally soft positive helicity gravitons give rise to $w_{1+\infty}$ and the negative helicity conformally soft gravitons give rise to an infinite Abelian algebra. The complete symmetry algebra is the semidirect product of the $w_{1+\infty}$ and the infinite Abelian algebra. This infinite Abelian algebra has not been discussed in the literature and we will show that this fills up a gap in the analysis of MHV amplitudes using Celestial holography.

\section{Knizhnik-Zamolodchikov type null states for negative helicity gravitons}\label{kz1}
The symmetry algebra \eqref{eq:w3} admits null states. In this section we are interested in the Knizhnik-Zamolodchikov (KZ) type null states which contain the $L_{-1}$ descendent of the (hard) graviton operator on the celestial sphere. Such KZ type null states in the MHV sector were found in \cite{Banerjee:2020zlg}. However, if we use only the symmetry algebra generated by the positive helicity (conformally soft) gravitons then we get only $(n-2)$ KZ type equations for an $n$-point MHV amplitude. The $(n-2)$ null states contain the $L_{-1}$ descendant of the \textit{positive} helicity hard graviotn only. So the list of KZ type null states obtained in \cite{Banerjee:2020zlg} was \textit{incomplete}. In this section we fill in this gap and find the two missing null states using the infinite Abelian symmetry algebra generated by the conformally soft negative helicity gravitons in the MHV sector. 

We start with the following mixed helicity OPE between two graviton primary operators in the MHV-sector up to $\mathcal{O}(\bar{z})$ \cite{Banerjee:2020zlg}
\begin{equation}
\label{gr_pm}
\begin{split}
G^+_{\Delta_1}\left(z_1,\bar{z}_1\right)G^-_{\Delta_2}\left(z_2,\bar{z}_2\right)=&\frac{\bar{z}_{12}}{z_{12}}B\left(\Delta_1-1,\Delta_2+3\right)H^1_{-\frac{1}{2},-\frac{1}{2}}G^-_{\Delta_1+\Delta_2-1}\left(z_2,\bar{z}_2\right)\\
&+B\left(\Delta_1-1,\Delta_2+3\right)H^1_{-\frac{3}{2},\frac{1}{2}}G^-_{\Delta_1+\Delta_2-1}\left(z_2,\bar{z}_2\right)\\
&+\bar{z}_{12}B\left(\Delta_1-1,\Delta_2+3\right)\bigg[\frac{\Delta_1-1}{\Delta_1+\Delta_2+2}H^0_{-1,0}G^-_{\Delta_1+\Delta_2}(z_2,\bar{z}_2)\\
&+\Delta_1~H^1_{-\frac{3}{2},-\frac{1}{2}}G^-_{\Delta_1+\Delta_2-1}(z_2,\bar{z}_2)\bigg]+\cdots
\end{split}
\end{equation}
Now taking $\Delta_2\to -3$ soft limit in \eqref{gr_pm} we get
\begin{equation}\label{ns}
\begin{split}
G^+_{\Delta_1}(z_1,\bar{z}_1)\bar{H}^{-3}(z_2,\bar{z}_2)=&-\frac{\bar{z}_{12}}{z_{12}}G^-_{\Delta_1-3}\left(z_2,\bar{z}_2\right)+H^1_{-\frac{3}{2},\frac{1}{2}}G^-_{\Delta_1-4}\left(z_2,\bar{z}_2\right)\\
&+\bar{z}_{12}\left[H^0_{-1,0}G^-_{\Delta_1-3}(z_2,\bar{z}_2)+\Delta_1~H^1_{-\frac{3}{2},-\frac{1}{2}}G^-_{\Delta_1-4}(z_2,\bar{z}_2)\right]+\cdots
\end{split}
\end{equation}
Now we expand the RHS of \eqref{ns} around $(z_1,\bar z_1)$
$$z_2\to z_1-z_{12},~~\bar{z}_2\to \bar{z}_{1}-\bar{z}_{12}~,$$  and demand consistency with the OPE between $\bar{H}^{-3}(z_2,\bar z_2)$ and $G_{\D}^+(z_1,\bar z_1)$. This leads to the following identity
\begin{equation}
\begin{split}
-\bar{z}_{12}\bar{H}^{-3}_{\frac{5}{2},-\frac{1}{2}}G^+_{\Delta_1}\left(z_1,\bar{z}_1\right)=&\bar{z}_{12}\bigg[L_{-1}G^-_{\Delta_1-3}(z_1,\bar{z}_1)+H^0_{0,-1}H^1_{-\frac{3}{2},\frac{1}{2}}G^-_{\Delta_1-4}(z_1,\bar{z}_1)\\
&+H^0_{-1,0}G^-_{\Delta_1-3}(z_1,\bar{z}_1)+\Delta_1~H^1_{-\frac{3}{2},-\frac{1}{2}}G^-_{\Delta_1-4}(z_1,\bar{z}_1)\bigg]\\
&
\end{split}
\end{equation}
which is equivalent to the existence of the following null state involving the $L_{-1}$ descendant of the negative helicity graviton
\begin{equation}
\begin{split}
\boxed{
L_{-1}G^-_{\Delta}+H^0_{-1,0}G^-_{\Delta}+\left(\Delta+3\right)H^1_{-\frac{3}{2},-\frac{1}{2}}G^-_{\Delta-1}+H^0_{0,-1}H^1_{-\frac{3}{2},\frac{1}{2}}G^-_{\Delta-1}+\bar{H}^{-3}_{\frac{5}{2},-\frac{1}{2}}G^+_{\Delta+3}=0}
\end{split}
\end{equation}
 
\section{Symmetry algebra from negative helicity gluon in the MHV-sector} \label{gluon}
Here we discuss the additional symmetry generators arising out of conformally soft negative helicity gluons in the MHV-sector. Our analysis is similar to the gravity case as discussed above. We denote $O^{a,\pm}_{\Delta}(z,\bar{z})$ as a positive/negative helicity outgoing gluon conformal primary operator of dimension $\Delta$ and adjoint group index $a$ at the point $(z,\bar{z})$ on the celestial torus. The singular terms of the OPE between a negative and a positive-helicity gluon conformal primary operators in the MHV-sector are given by \cite{Guevara:2021abz}
\begin{equation}
	\label{eq:gope1}
\begin{split}
O^{a,-}_{\Delta_1}(z_1,\bar z_1) O^{b,+}_{\Delta_2}(z_2,\bar z_2) \sim 
-\frac{if^{ab}_{~~c}}{z_{12}} \sum_{n=0}^{\infty} B(\Delta_1 +1 +n, \Delta_2 -1) \frac{(\bar z_{12})^n}{n!} \bar\partial_2^n O^{c,-}_{\Delta_1+\Delta_2-1}(z_2,\bar z_2).
\end{split}
\end{equation}
Now we define the "conformally soft" negative-helicity gluon operators $\bar R^{k,a}(z,\bar z)$ as
\begin{equation}
\begin{split}
\bar R^{k,a}(z,\bar z) := \lim_{\Delta\rightarrow k} (\Delta-k) O^{a,-}_{\Delta}(z,\bar z),~~~k=-1,-2,-3, \cdots
\end{split}
\end{equation}
with weights $$\left(h,\bar{h}\right)=\left(\frac{k-1}{2},\frac{k+1}{2}\right)~~.$$
The structure of OPE \eqref{eq:gope1} allows us to do the following truncated mode expansion of $\bar R^{k,a}(z,\bar z)$ in $\bar{z}$-variable
\begin{equation}
\begin{split}
\bar R^{k,a}(z,\bar z) = \sum_{m= \frac{k+1}{2}}^{-\frac{k+1}{2}} \frac{\bar R^{k,a}_m(z)}{\bar z^{m+\frac{k+1}{2}}}
\end{split}
\end{equation}
We can further mode expand the holomorphic currents $\bar R^{k,a}_m(z)$ in $z$-variable 
\be
\bar R^{k,a}_m(z) = \sum_{\alpha \in \mathbb{Z} - \frac{k-1}{2}} \frac{\bar R^{k,a}_{\alpha,m}}{z^{\alpha+\frac{k-1}{2}}}
\ee
Similarly, we define an infinite tower of conformally soft positive-helicity gluon operators  
\begin{equation}
	\begin{split}
		R^{k,a}\left(z,\bar{z}\right)=\lim_{\Delta\to k}\left(\Delta-k\right)O^{a,+}_{\Delta}\left(z,\bar{z}\right),~~~k=1,0,-1,\cdots
	\end{split}
\end{equation}
with weights $$\left(h,\bar{h}\right)=\left(\frac{k+1}{2},\frac{k-1}{2}\right)~~.$$
The modes of $R^{k,a}(z,\bar{z})$ satisfy the following holographic symmetry algebra \cite{Guevara:2021abz}
\begin{equation}
\begin{split}
\left[R^{k,a}_{\alpha,n},{R}^{l,b}_{\alpha^{\prime},n^{\prime}}\right]=-if^{ab}_{~~c}~\frac{\left(\frac{1-k}{2}-n+\frac{1-l}{2}-n^{\prime}\right)!\left(\frac{1-k}{2}+n+\frac{1-l}{2}+n^{\prime}\right)!}{\left(\frac{1-k}{2}-n\right)!\left(\frac{1-l}{2}-n^{\prime}\right)!\left(\frac{1-k}{2}+n\right)!\left(\frac{1-l}{2}+n^{\prime}\right)!}{R}^{k+l-1,c}_{\alpha+\alpha^{\prime}, n+n^{\prime}}
\end{split}
\end{equation}
Similarly, we obtain the following commutator algebra between the modes of $R^{k,a}(z,\bar{z})$ and $\bar{R}^{l,b}(z,\bar{z})$ from the mixed helicity OPE \eqref{eq:gope1}(the detail of the computation is in Appendix \ref{gluon_b})
\begin{equation}
\begin{split}
\left[R^{k,a}_{\alpha,n},\bar{R}^{l,b}_{\alpha^{\prime},n^{\prime}}\right]=-if^{ab}_{~~c}~\frac{\left(\frac{1-k}{2}-n-\frac{l+1}{2}-n^{\prime}\right)!\left(\frac{1-k}{2}+n-\frac{l+1}{2}+n^{\prime}\right)!}{\left(\frac{1-k}{2}-n\right)!\left(-\frac{l+1}{2}-n^{\prime}\right)!\left(\frac{1-k}{2}+n\right)!\left(-\frac{l+1}{2}+n^{\prime}\right)!}\bar{R}^{k+l-1,c}_{\alpha+\alpha^{\prime}, n+n^{\prime}}
\end{split}
\end{equation}
The OPE between two negative helicity gluons does not a have pole term, so we get
\begin{equation}
\begin{split}
\left[\bar{R}^{k,a}_{\alpha,n},\bar{R}^{l,b}_{\alpha^{\prime},n^{\prime}}\right]=0.
\end{split}
\end{equation}
Now we define the conformally soft light transformed positive and negative helicity gluon operators as \cite{Strominger:2021lvk}
\begin{equation}
	\label{eq:gluon1}
	S^{p,a}_{m}=(p-1-m)!(p-1+m)!R^{3-2p,a}_{m}
\end{equation}
\begin{equation}
	\label{eq:gluon2}
	\bar{S}^{q,a}_{n}=(q-1-n)!(q-1+n)!\bar{R}^{1-2q,a}_{m}
\end{equation}
where $p,q=1,\frac{3}{2},2,\frac{5}{2},...$  
with the following restrictions on $m$ and $n$
\begin{equation}
	\label{eq:res2}
	1-p \le m \le p-1; \hspace{4mm}  1-q \le n \le q-1.
\end{equation}
The algebra of these light transformed operators is given by
\begin{equation}
\label{eq:gluon3}
[S^{p,a}_{m},S^{q,b}_{n}]=-i{f}^{ab}_{~~c}S^{p+q-1,c}_{m+n}
\end{equation}
\begin{equation}
	\label{eq:gluon4}
	[S^{p,a}_{m},\bar{S}^{q,b}_{n}]=-i{f}^{ab}_{~~c}\bar{S}^{p+q-1,c}_{m+n}
\end{equation}
\begin{equation}
	\label{eq:gluon5}
	[\bar{S}^{p,a}_{m},\bar{S}^{q,b}_{n}]=0
\end{equation}
Maximally helicity violating(MHV)-gluon scattering amplitudes in Minkowski spacetime are governed by the above symmetry algebra. The conformally soft positive helicity gluons give rise to the S-algebra \eqref{eq:gluon3} and the conformally soft negative helicity gluons give rise to an infinite Abelian algebra \eqref{eq:gluon5}. The complete symmetry algebra of MHV-gluon scattering amplitudes is the semi-direct product of the S-algebra and the infinite Abelian algebra.
\section{Knizhnik-Zamolodchikov type null states for \textit{negative} helicity gluons}\label{kz2}
In this section we obtain the KZ type null state for the negative helicity gluon using the additional symmetries discussed in section \ref{gluon}. We proceed in the same way as in the case of gravitons. The mixed helicity OPE between two gluon primary operators up to $\mathcal{O}(1)$ is given by \cite{Banerjee:2020vnt}
\begin{equation}
\label{eq:gkz1}
\begin{split}
O^{a,+}_{\Delta_1}(z_1,\bar{z}_1)O^{b,-}_{\Delta_2}(z_2,\bar{z}_2)=&B\left(\Delta_1-1,\Delta_2+1\right)\bigg[-\frac{if^{ab}_{~~c}}{z_{12}}+\Delta_1\delta^{bc}R^{1,a}_{-1,0}\\
&+\frac{\Delta_1-1}{\Delta_1+\Delta_2}\delta^{bc}R^{0,a}_{-\frac{1}{2},\frac{1}{2}}\left(-H^1_{-\frac{1}{2},-\frac{1}{2}}\right)\bigg]O^{c,-}_{\Delta_1+\Delta_2-1}(z_2,\bar{z}_2)
\end{split}
\end{equation}
Now we take the $\D_2\rightarrow -1$ soft limit in \eqref{eq:gkz1}. Then we demand the consistency with the OPE between $\bar{R}^{-1,b}$ and $O^{a,+}_{\Delta_1}$ and comparing the $\mathcal{O}(1)$ terms, we obtain the following null state relations involving negative helicity gluon (the details of the derivation is in appendix \ref{kz_d}) 
\begin{equation}
\begin{split}
\boxed{
C_AL_{-1}O^{a,-}_{\Delta}-(\Delta+2)R^{1,b}_{-1,0}R^{1,b}_{0,0}O^{a,-}_{\Delta}-R^{0,b}_{-\frac{1}{2},\frac{1}{2}}R^{1,b}_{0,0}O^{a,-}_{\Delta+1}-\bar{R}^{-1,b}_{1,0}R^{1,b}_{0,0}O^{a,+}_{\Delta+2}=0.}
\end{split}
\end{equation}
where $C_A$ is the quadratic Casimir of the adjoint representation.

\section{Acknowledgement}
The work of SB is supported by the  Swarnajayanti Fellowship (File No- SB/SJF/2021-22/14) of the Department of Science and Technology and ANRF, India. The work of MP is supported by an IOE endowed Postdoctoral position at IISc, Bengaluru, India.

\appendix
\section{Algebra for graviton}\label{graviton_a}
Let us start with the mixed helicity OPE for gravitons in the MHV sector
\begin{equation}
	\label{eq:ap1}
\begin{split}
G^+_{\Delta_1}(z_1,\bar z_1) G^-_{\Delta_2}(z_2,\bar z_2) \sim 
-\frac{\bar z_{12}}{z_{12}} \sum_{n=0}^{\infty} B(\Delta_1 -1 +n, \Delta_2 +3) \frac{(\bar z_{12})^n}{n!} \bar\partial_2^n G^-_{\Delta_1+\Delta_2}(z_2,\bar z_2).
\end{split}
\end{equation}
Now we define the "conformally soft" negative-helicity graviton operator $\bar H^k(z,\bar z)$ as
\begin{equation}
\begin{split}
\bar H^k(z,\bar z) = \lim_{\Delta\rightarrow k} (\Delta-k) G^-_{\Delta}(z,\bar z),~~~k=-3,-4,-5,...
\end{split}
\end{equation}
\eqref{eq:ap1} allows us to do the following expansion of $\bar H^k(z,\bar z)$
\begin{equation}
\begin{split}
\bar H^k(z,\bar z) = \sum_{m= \frac{k+2}{2}}^{-\frac{k+2}{2}} \frac{\bar H^k_m(z)}{\bar z^{m+\frac{k+2}{2}}}
\end{split}
\end{equation}
Similarly, one can define conformally soft positive-helicity graviton as
\begin{equation}
\begin{split}
H^k\left(z,\bar{z}\right)=\lim_{\Delta\to k}\left(\Delta-k\right)G^+_{\Delta}\left(z,\bar{z}\right),~~~k=1,0,-1,\cdots
\end{split}
\end{equation}
with weights $\left(\frac{k+2}{2},\frac{k-2}{2}\right)$.\\
$H^k\left(z,\bar{z}\right)$ also has the following truncated mode expansion
\begin{equation}
\begin{split}
H^k\left(z,\bar{z}\right)=\sum_{m=\frac{k-2}{2}}^{\frac{2-k}{2}}\frac{H^k_m(z)}{\bar{z}^{m+\frac{k-2}{2}}}
\end{split}
\end{equation}
Taking the $\Delta_1$ soft in \eqref{eq:ap1} we get
\begin{equation}
	\label{eq:ap2}
\begin{split}
H^k(z_1,\bar z_1) G^-_{\Delta_2}(z_2,\bar z_2) \sim 
-\frac{\bar z_{12}}{z_{12}} \sum_{n=0}^{1-k} \frac{(-1)^{1-k-n}}{\left(1-k-n\right)!}\frac{\Gamma\left(\Delta_2+3\right)}{\Gamma\left(k+\Delta_2+n+2\right)} \frac{(\bar z_{12})^n}{n!} \bar\partial_2^n G^-_{k+\Delta_2}(z_2,\bar z_2)
\end{split}
\end{equation}
Now, making $\Delta_2$ soft in \eqref{eq:ap2} leads to the following $H\bar{H}$ OPE
\begin{equation}
\label{OPEHbH}
\begin{split}
H^k(z_1,\bar z_1) \bar H^l(z_2,\bar z_2) \sim 
-\frac{\bar z_{12}}{z_{12}} \sum_{n=0}^{1-k} \frac{\left(-k-l-n-2\right)!}{\left(-l-3\right)!\left(1-k-n\right)!} \frac{(\bar z_{12})^n}{n!} \bar\partial_2^n \bar H^{k+l}(z_2,\bar z_2)
\end{split}
\end{equation}
To determine the algebra of the modes, we first recall that the modes are extracted from
\begin{equation}
\begin{split}
H^k_n(z)=\oint \frac{d\bar{z}}{2\pi i}\bar{z}^{n+\frac{k-4}{2}}H^k\left(z,\bar{z}\right)
\end{split}
\end{equation}
and
\begin{equation}
\begin{split}
\bar{H}^l_{n^{\prime}}(z)=\oint \frac{d\bar{z}}{2\pi i}\bar{z}^{{n^{\prime}}+\frac{l}{2}}\bar{H}^l\left(z,\bar{z}\right)
\end{split}
\end{equation}
Then the algebra of the modes is obtained from the following integral
\begin{equation}
\begin{split}
\left[H^k_n,\bar{H}^l_{n^{\prime}}\right](z_2)=\oint_{|\bar{z}_1|<\epsilon}\frac{d\bar{z}_1}{2\pi i}\bar{z}_1^{n+\frac{k-4}{2}}\oint_{|\bar{z}_2|<\epsilon}\frac{d\bar{z}_2}{2\pi i}\bar{z}_2^{n^{\prime}+\frac{l}{2}}\oint_{|{z}_{12}|<\epsilon}\frac{d{z}_{1}}{2\pi i}H^k(z_1,\bar z_1) \bar H^l(z_2,\bar z_2)
\end{split}
\end{equation}
Since the OPE in \eqref{OPEHbH} is not singular in the antiholomorphic variables, the order in which the contour integrals in the antiholomorphic variables is taken does not matter.\\
Now we substitute \eqref{OPEHbH} in the r.h.s. Then we first perform the $z_1$ integral and use the following
\begin{equation}
\begin{split}
\oint_{|\bar{z}_1|<\epsilon}\frac{d\bar{z}_1}{2\pi i}\bar{z}_1^{n+\frac{k-4}{2}}\left(\bar{z}_{12}\right)^{m+1}&=0,~~~~~~~~~~~~~~~~~~~~~~~~~~~~~~~~~~~~~~~~~~~~~~~~~-1\leq m <-n-\frac{k}{2}\\
&=\frac{1}{\left(1-\frac{k}{2}-n\right)!}\frac{\left(m+1\right)!}{\left(m+n+\frac{k}{2}\right)!}\left(-\bar{z}_2\right)^{m+n+\frac{k}{2}},~~~~~~~-n-\frac{k}{2}\leq m\leq 1-k
\end{split}
\end{equation}
Then the commutator becomes
\begin{equation}
\begin{split}
\left[H^k_n,\bar{H}^l_{n^{\prime}}\right](z_2)=-\sum_{m=-n-\frac{k}{2}}^{1-k}&\frac{\left(-k-l-m-2\right)!}{\left(-l-3\right)!\left(1-k-m\right)!}\frac{(m+1)\left(-1\right)^{m+n+\frac{k}{2}}}{\left(1-\frac{k}{2}-n\right)!\left(m+n+\frac{k}{2}\right)!}\\
&\oint_{|\bar{z}_2|<\epsilon}\frac{d\bar{z}_2}{2\pi i}\left(\bar{z}_2\right)^{m+n+\frac{k}{2}+n^{\prime}+\frac{l}{2}}\partial_{\bar{z}_2}^m\bar{H}^{k+l}\left(z_2,\bar{z}_2\right)
\end{split}
\end{equation}
To perform the remaining contour integral we substitute $\bar{H}^{k+l}$ for its mode expansion
\begin{equation}
\nonumber
\begin{split}
\oint_{|\bar{z}_2|<\epsilon}\frac{d\bar{z}_2}{2\pi i}\left(\bar{z}_2\right)^{m+n+\frac{k}{2}+n^{\prime}+\frac{l}{2}}\partial_{\bar{z}_2}^m\bar{H}^{k+l}\left(z_2,\bar{z}_2\right)&=\oint_{|\bar{z}_2|<\epsilon}\frac{d\bar{z}_2}{2\pi i}\left(\bar{z}_2\right)^{m+n+\frac{k}{2}+n^{\prime}+\frac{l}{2}}\partial_{\bar{z}_2}^m\sum_{m^{\prime}=\frac{k+l+2}{2}}^{-\frac{k+l+2}{2}}\frac{\bar{H}^{k+l}_{m^{\prime}}(z_2)}{\bar{z}_2^{m^{\prime}+\frac{k+l+2}{2}}}\\
&=\frac{\left(-n-n^{\prime}-\frac{k+l+2}{2}\right)!}{\left(-n-n^{\prime}-\frac{k+l+2}{2}-m\right)!}\bar{H}^{k+l}_{n+n^{\prime}}(z_2)
\end{split}
\end{equation}
Substituting the result back and performing the sum we have
\begin{equation}
\begin{split}
\left[H^k_n,\bar{H}^l_{n^{\prime}}\right](z_2)=&-\left[n^{\prime}(2-k)+n(2+l)\right]\\
&\times\frac{\left(\frac{2-k}{2}-n-\frac{2+l}{2}-n^{\prime}-1\right)!}{\left(\frac{2-k}{2}-n\right)!\left(-\frac{2+l}{2}-n^{\prime}\right)!}\frac{\left(\frac{2-k}{2}+n-\frac{2+l}{2}+n^{\prime}-1\right)!}{\left(\frac{2-k}{2}+n\right)!\left(-\frac{2+l}{2}+n^{\prime}\right)!}\bar{H}^{k+l}_{n+n^{\prime}}(z_2)
\end{split}
\end{equation}
This is the conformal soft graviton algebra.
\section{Algebra for gluon}\label{gluon_b}
We start with the mixed helicity OPE for the gluon.
\begin{equation}
\begin{split}
O^{a,+}_{\Delta_1}(z_1,\bar z_1) O^{b,-}_{\Delta_2}(z_2,\bar z_2) \sim 
-\frac{if^{abc}}{z_{12}} \sum_{m=0}^{\infty} B(\Delta_1 -1 +m, \Delta_2 +1) \frac{(\bar z_{12})^m}{m!} \bar\partial_2^m O^{c,-}_{\Delta_1+\Delta_2-1}(z_2,\bar z_2)
\end{split}
\end{equation}
Now we define the "soft" operator $\bar R^{k,a}(z,\bar z)$ as
\begin{equation}
\begin{split}
\bar R^{k,a}(z,\bar z) := \lim_{\Delta\rightarrow k} (\Delta-k) O^{a,-}_{\Delta}(z,\bar z)
\end{split}
\end{equation}
So we do the following expansion of $\bar R^{k,a}(z,\bar z)$
\begin{equation}
\begin{split}
\bar R^{k,a}(z,\bar z) = \sum_{m= \frac{k+1}{2}}^{-\frac{k+1}{2}} \frac{\bar R^{k,a}_m(z)}{\bar z^{m+\frac{k+1}{2}}}
\end{split}
\end{equation}
Similarly, we define
\begin{equation}
\begin{split}
R^{k,a}\left(z,\bar{z}\right)=\lim_{\Delta\to k}\left(\Delta-k\right)O^{a,+}_{\Delta}\left(z,\bar{z}\right),~~~k=1,0,-1,\cdots
\end{split}
\end{equation}
Similarly, we have the following truncated mode expansion
\begin{equation}
\begin{split}
R^{k,a}\left(z,\bar{z}\right)=\sum_{m=\frac{k-1}{2}}^{\frac{1-k}{2}}\frac{R^{k,a}_m(z)}{\bar{z}^{m+\frac{k-1}{2}}}
\end{split}
\end{equation}
After taking the $\Delta_1$ soft we have
\begin{equation}
\begin{split}
R^{k,a}(z_1,\bar z_1) O^{b,-}_{\Delta_2}(z_2,\bar z_2) \sim 
-\frac{if^{ab}_{~~c}}{z_{12}} \sum_{m=0}^{1-k} \frac{(-1)^{1-k-m}}{\left(1-k-m\right)!}\frac{\Gamma\left(\Delta_2+1\right)}{\Gamma\left(k+\Delta_2+m\right)} \frac{(\bar z_{12})^m}{m!} \partial_{\bar{z}_2}^m O^{c,-}_{k+\Delta_2-1}(z_2,\bar z_2)
\end{split}
\end{equation}
Now, after taking $\Delta_2$ soft we have
\begin{equation}
\label{OPEHbR}
\begin{split}
R^{k,a}(z_1,\bar z_1) \bar R^{l,b}(z_2,\bar z_2) \sim 
-\frac{if^{ab}_{~~c}}{z_{12}} \sum_{m=0}^{1-k} \frac{\left(-k-l-m\right)!}{\left(-l-1\right)!\left(1-k-m\right)!} \frac{(\bar z_{12})^m}{m!} \partial_{\bar{z}_2}^m \bar R^{k+l-1,c}(z_2,\bar z_2)
\end{split}
\end{equation}
To determine the algebra of the modes, we first recall that the modes are extracted from
\begin{equation}
\begin{split}
R^{k,a}_n(z)=\oint \frac{d\bar{z}}{2\pi i}\bar{z}^{n+\frac{k-3}{2}}R^{k,a}\left(z,\bar{z}\right)
\end{split}
\end{equation}
and
\begin{equation}
\begin{split}
\bar{R}^{l,a}_{n^{\prime}}(z)=\oint \frac{d\bar{z}}{2\pi i}\bar{z}^{{n^{\prime}}+\frac{l-1}{2}}\bar{R}^{l,a}\left(z,\bar{z}\right)
\end{split}
\end{equation}
Then the algebra of the modes is obtained from the following integral
\begin{equation}
\begin{split}
\left[R^{k,a}_n,\bar{R}^{l,b}_{n^{\prime}}\right](z_2)=\oint_{|\bar{z}_1|<\epsilon}\frac{d\bar{z}_1}{2\pi i}\bar{z}_1^{n+\frac{k-3}{2}}\oint_{|\bar{z}_2|<\epsilon}\frac{d\bar{z}_2}{2\pi i}\bar{z}_2^{n^{\prime}+\frac{l-1}{2}}\oint_{|{z}_{12}|<\epsilon}\frac{d{z}_{1}}{2\pi i}R^{k,a}(z_1,\bar z_1) \bar R^{l,b}(z_2,\bar z_2)
\end{split}
\end{equation}
Here again, since the OPE in \eqref{OPEHbR} is not singular in the antiholomorphic variables, the order in which the contour integrals in the antiholomorphic variables is taken does not matter.\\
Now we substitute \eqref{OPEHbR} in the r.h.s. Then we first perform the $z_1$ integral and use the following
\begin{equation}
\begin{split}
\oint_{|\bar{z}_1|<\epsilon}\frac{d\bar{z}_1}{2\pi i}\bar{z}_1^{n+\frac{k-3}{2}}\left(\bar{z}_{12}\right)^{m}&=0,~~~~~~~~~~~~~~~~~~~~~~~~~~~~~~~~~~~~~~~~~~~~~~~~0\leq m <\frac{1-k}{2}-n\\
&=\frac{m!}{\left(\frac{1-k}{2}-n\right)!\left(m+n+\frac{k-1}{2}\right)!}\left(-\bar{z}_2\right)^{m+n+\frac{k-1}{2}},~~~~~~~\frac{1-k}{2}-n\leq m\leq 1-k
\end{split}
\end{equation}
Then the commutator becomes
\begin{equation}
\begin{split}
\left[R^{k,a}_n,\bar{R}^{l,b}_{n^{\prime}}\right](z_2)=&-if^{ab}_{~~c}\sum_{m=\frac{1-k}{2}-n}^{1-k}\frac{\left(-k-l-m\right)!}{\left(-l-1\right)!\left(1-k-m\right)!}\frac{\left(-1\right)^{m+n+\frac{k-1}{2}}}{\left(\frac{1-k}{2}-n\right)!\left(m+n+\frac{k-1}{2}\right)!}\\
&\times \oint_{|\bar{z}_2|<\epsilon}\frac{d\bar{z}_2}{2\pi i}\bar{z}_2^{m+n+\frac{k-1}{2}+n^{\prime}+\frac{l-1}{2}}\partial_{\bar{z}_2}^m\bar{R}^{k+l-1,c}\left(z_2,\bar{z}_2\right)
\end{split}
\end{equation}
To perform the remaining contour integral we substitute $\bar{R}^{k+l-1,c}$ for its mode expansion
\begin{equation}
\begin{split}
\oint_{|\bar{z}_2|<\epsilon}\frac{d\bar{z}_2}{2\pi i}\bar{z}_2^{m+n+\frac{k-1}{2}+n^{\prime}+\frac{l-1}{2}}\partial_{\bar{z}_2}^m\bar{R}^{k+l-1,c}\left(z_2,\bar{z}_2\right)&=\oint_{|\bar{z}_2|<\epsilon}\frac{d\bar{z}_2}{2\pi i}\bar{z}_2^{m+n+\frac{k-1}{2}+n^{\prime}+\frac{l-1}{2}}\partial_{\bar{z}_2}^m\sum_{m^{\prime}=\frac{k+l}{2}}^{-\frac{k+l}{2}}\frac{\bar{R}^{k+l-1,c}_{m^{\prime}}(z_2)}{\bar{z}_2^{m^{\prime}+\frac{k+l}{2}}}\\
&=\frac{\left(-n-n^{\prime}-\frac{k+l}{2}\right)!}{\left(-n-n^{\prime}-m-\frac{k+l}{2}\right)!}\bar{R}^{k+l-1,c}_{n+n^{\prime}}(z_2)
\end{split}
\end{equation}
Substituting the result back and performing the sum we have
\begin{equation}
\begin{split}
\left[R^{k,a}_n,\bar{R}^{l,b}_{n^{\prime}}\right](z_2)=-if^{ab}_{~~c}~\frac{\left(\frac{1-k}{2}-n-\frac{l+1}{2}-n^{\prime}\right)!\left(\frac{1-k}{2}+n-\frac{l+1}{2}+n^{\prime}\right)!}{\left(\frac{1-k}{2}-n\right)!\left(-\frac{l+1}{2}-n^{\prime}\right)!\left(\frac{1-k}{2}+n\right)!\left(-\frac{l+1}{2}+n^{\prime}\right)!}\bar{R}^{k+l-1,c}_{n+n^{\prime}}(z_2)
\end{split}
\end{equation}
This is the conformal soft gluon algebra.
%
%
\section{The $KZ$ type null state for gluon}\label{kz_d}
In this appendix, we derive the Knizhnik-Zamolodchikov (KZ) type null state for negaive-helicity gluon operators. These null states are obtained by taking the soft limits in the OPE and demanding the consistency with the OPE between $\bar{R}$ and $O^{a,+}$. The OPE between a positive and a negative-helicity gluon primary operators up to $\mathcal{O}(1)$ is given by \cite{Banerjee:2020vnt}
\begin{equation}
\label{gl_a}
\begin{split}
O^{a,+}_{\Delta_1}(z_1,\bar{z}_1)O^{b,-}_{\Delta_2}(z_2,\bar{z}_2)=&B\left(\Delta_1-1,\Delta_2+1\right)\bigg[-\frac{if^{ab}_{~~c}}{z_{12}}+\Delta_1\delta^{bc}R^{1,a}_{-1,0}\\
&+\frac{\Delta_1-1}{\Delta_1+\Delta_2}\delta^{bc}R^{0,a}_{-\frac{1}{2},\frac{1}{2}}\left(-H^1_{-\frac{1}{2},-\frac{1}{2}}\right)\bigg]O^{c,-}_{\Delta_1+\Delta_2-1}(z_2,\bar{z}_2)
\end{split}
\end{equation}
And then we take $\Delta_2\to -1$ soft limit on both side of \eqref{gl_a}. Then we have
\begin{equation}
\label{gl-int}
\begin{split}
O^{a,+}_{\Delta_1}(z_1,\bar{z}_1)\bar{R}^{-1,b}(z_2,\bar{z}_2)&=\bigg[-\frac{if^{ab}_{~~c}}{z_{12}}+\Delta_1\delta^{bc}R^{1,a}_{-1,0}+\delta^{bc}R^{0,a}_{-\frac{1}{2},\frac{1}{2}}\left(-H^1_{-\frac{1}{2},-\frac{1}{2}}\right)\bigg]O^{c,-}_{\Delta_1-2}(z_2,\bar{z}_2)
\end{split}
\end{equation}
Now we demand the consistency with the OPE between $\bar{R}^{-1,b}(z_2,\bar{z}_2)$ and $O^{a,+}_{\Delta_1}(z_1,\bar{z}_1)$ and expand around by
$$z_2\to z_1-z_{12},~~\bar{z}_2\to \bar{z}_{1}-\bar{z}_{12}~.$$ After comparing the $\mathcal{O}(1)$ terms, we have
\begin{equation}
\begin{split}
\bar{R}^{-1,b}_{1,0}O^{a,+}_{\Delta_1}(z_1,\bar{z}_1)=if^{ab}_{~c}L_{-1}O^{c,-}_{\Delta_1-2}(z_1,\bar{z}_1)+\Delta_1R^{1,a}_{-1,0}
\end{split}O^{b,-}_{\Delta_1-2}(z_1,\bar{z}_1)+R^{0,a}_{-\frac{1}{2},\frac{1}{2}}O^{b,-}_{\Delta_1-1}(z_1,\bar{z}_1)
\end{equation}
Hence, we obtain the new KZ type null states involving negative helicity gluons in the MHV-sector
\begin{equation}
\begin{split}
if^{ab}_{~~c}~L_{-1}O^{c,-}_{\Delta}+(\Delta+2)R^{1,a}_{-1,0}O^{b,-1}_{\Delta}+R^{0,a}_{-\frac{1}{2},\frac{1}{2}}O^{b,-}_{\Delta+1}-\bar{R}^{-1,b}_{1,0}O^{a,+}_{\Delta+2}=0.
\end{split}
\end{equation}
Multiplying with $-if^{abd}$ and using $f^{aa_1b}f^{aa_1c}=C_A\delta^{bc}$, we will get the following relations
\begin{equation}
\begin{split}
C_AL_{-1}O^{a,-}_{\Delta}-(\Delta+2)R^{1,b}_{-1,0}R^{1,b}_{0,0}O^{a,-}_{\Delta}-R^{0,b}_{-\frac{1}{2},\frac{1}{2}}R^{1,b}_{0,0}O^{a,-}_{\Delta+1}-\bar{R}^{-1,b}_{1,0}R^{1,b}_{0,0}O^{a,+}_{\Delta+2}=0.
\end{split}
\end{equation}
where $C_A$ is the quadratic Casimir of the adjoint representation.
%
%
%
%
%
\section{Graviton primaries of the new symmetry algebra} \label{primaries}
In this Appendix, we write down the conditions on the primary operators that follow from the OPE between two outgoing mixed helicity graviton primaries. The detail procedure of obtaining these conditions is in Appendix F of \cite{Banerjee:2023jne}.
We have the following conditions
\begin{equation}
\begin{split}
H^k_{-\frac{k+2}{2}+m,\frac{2-k}{2}-n-1}G^-_{\Delta}\left(0,0\right)=-\frac{(-1)^{1-k-n}}{\left(1-k-n\right)!}\frac{\Gamma\left(\Delta+3\right)}{\Gamma\left(k+\Delta+n+2\right)}\frac{1}{n!}\bar{\partial}^nG^-_{\Delta+k}(0,0)
\end{split}
\end{equation}
for $m=1$ and $0\leq n\leq 1-k$ and $k=1, 0, -1, \cdots$
and
\begin{equation}
\begin{split}
H^k_{-\frac{k+2}{2}+m,\frac{2-k}{2}-n-1}G^-_{\Delta}\left(0,0\right)=0
\end{split}
\end{equation}
for $m>1$ and $0\leq n\leq 1-k$ and $k=1, 0, -1, \cdots$\\
And we also have
\begin{equation}
\begin{split}
H^k_{-\frac{k+2}{2}+m,\frac{2-k}{2}}G^-_{\Delta}\left(0,0\right)=0
\end{split}
\end{equation}
for $m\geq 1$ and $k=1, 0, -1, \cdots$~.\\\

Similarly, we have the following conditions
\begin{equation}
\begin{split}
\bar{H}^k_{m-\frac{k-2}{2},-\frac{k+2}{2}-n-1}G^+_{\Delta}(0,0)=-\frac{(-1)^{-k-n-3}}{\left(-k-n-3\right)!}\frac{\Gamma\left(\Delta-1\right)}{\Gamma\left(k+\Delta+n+2\right)}\frac{1}{n!}\bar{\partial}^nG^-_{\Delta+k}(0,0)
\end{split}
\end{equation}
for $m=1$ and $0\leq n\leq -k-3$~. and
\begin{equation}
\begin{split}
\bar{H}^k_{-\frac{k-2}{2}+m,-\frac{k+2}{2}-n-1}G^+_{\Delta}(0,0)=0
\end{split}
\end{equation}
for $m>1$ and $0\leq n\leq -k-3$~.\\\
And we also have
\begin{equation}
\begin{split}
\bar{H}^k_{-\frac{k-2}{2}+m,-\frac{k+2}{2}}G^+_{\Delta}\left(0,0\right)=0
\end{split}
\end{equation}
for $m\geq 1$~.\\\

From the OPE between two positive-helicity outgoing gravitons, we have the following conditions
\begin{equation}
\begin{split}
H^k_{-\frac{k+2}{2}+m,\frac{2-k}{2}-n-1}G^+_{\Delta}\left(0,0\right)=-\frac{(-1)^{1-k-n}}{\left(1-k-n\right)!}\frac{\Gamma\left(\Delta-1\right)}{\Gamma\left(k+\Delta+n-2\right)}\frac{1}{n!}\bar{\partial}^nG^+_{\Delta+k}(0,0)
\end{split}
\end{equation}
for $m=1$ and $0\leq n\leq 1-k$ and $k=1, 0, -1, \cdots$
and
\begin{equation}
\begin{split}
H^k_{-\frac{k+2}{2}+m,\frac{2-k}{2}-n-1}G^+_{\Delta}\left(0,0\right)=0
\end{split}
\end{equation}
for $m>1$ and $0\leq n\leq 1-k$ and $k=1, 0, -1, \cdots$\\
And we also have
\begin{equation}
\begin{split}
H^k_{-\frac{k+2}{2}+m,\frac{2-k}{2}}G^+_{\Delta}\left(0,0\right)=0
\end{split}
\end{equation}
for $m\geq 1$ and $k=1, 0, -1, \cdots$~.

\end{document}